\documentclass[a4paper,11pt,5p]{elsarticle}
\usepackage{siunitx}
\usepackage{multirow}		
\usepackage{ctable}			
\usepackage{threeparttable}
\usepackage{multicol, blindtext}
\usepackage{csquotes}		

\biboptions{sort&compress}	
\begin{document}

\begin{frontmatter}

	\title{Mechanisms of Elevated Temperature Galling in Hardfacings}

		\author[IC]{Samuel R. Rogers\corref{cor1}}
		\ead{srr13@ic.ac.uk}
		\author[RR]{David Stewart}
		\author[RR]{Paul Taplin}
		\author[IC]{David Dye}
		\cortext[cor1]{Corresponding author}
		\address[IC]{Imperial College, South Kensington, London SW7 2AZ, UK}
		\address[RR]{Rolls-Royce plc, Raynesway, Derby DE21 7WA, UK}

	\begin{abstract}
		The galling mechanism of Tristelle 5183, an Fe-based hardfacing alloy, was investigated at elevated temperature. The test was performed using a bespoke galling rig. Adhesive transfer and galling were found to occur, as a result of shear at the adhesion boundary and the activation of an internal shear plane within one of the tribosurfaces. During deformation, carbides were observed to have fractured, as a result of the shear train they were exposed to and their lack of ductility. In the case of niobium carbides, their fracture resulted in the formation of voids, which were found to coalesce and led to cracking and adhesive transfer. A tribologically affected zone (TAZ) was found to form, which contained nanocrystalline austenite, as a result of the shear exerted within \SI{30}{\micro\meter} of the adhesion boundaries. The galling of Tristelle 5183 initiated from the formation of an adhesive boundary, followed by sub-surface shear in only one tribosurface, Following further sub-surface shear, an internal shear plane is activated. internal shear and shear at the adhesion boundary continues until fracture occur, resulting in adhesive transfer.

	\end{abstract}
	
\end{frontmatter}

\section{Introduction}	
	Hardfacing alloys have been used for many years in components where surface degradation is of concern, particularly when coupled with extreme environments; be it high wear, high temperature, corrosive or erosive environments, or a combination of these. Of particular interest is the use of hardfacings in valve seatings within the pressurised water reactor (PWR) environment. This is a corrosive environment at \SI{300}{\degreeCelsius}, and in which wear and galling (severe adhesive wear) are of critical concern. The galling of components may result in gross surface degradation and component seizure, and is characterised by the plastic deformation and adhesion of contacting asperities \cite{TribologyBook}. The mobilisation of Co debris from \textit{e.g.} Stellite hardfacings is of concern due to the $\gamma$-radiation fields that are subsequently produced. As such, the reduction or replacement of cobalt alloys is necessary \cite{ONRCobalt}. Fe-based hardfacings are viewed as a potential replacement system, resulting in increased interest in the galling mechanisms, and with a number of alloys being designed and investigated \cite{SSFriction,RTOcken,GallingTorque,FrenchGalling,GallTough,GallingTemperature,416Galling,304Galling,Paper2,BowdenNature,ChongPaper,PooleGND,EPRINorem}. Galling has been found to occur most readily in systems which have limited movement perpendicular to the sliding direction, such as in valves \cite{SteelersGalling,Macdonald1971}. 
	
    Although work in the literature has investigated some of the mechanisms of galling, these have been focused upon single phase materials \cite{GallingBook,UnOxvsOx,Sasada1976,Cocks1962,Cocks1964,Antler1963,10YearStudy}, with little work investigating more complex microstructures, such as duplex microstructures or microstructures that include carbides or other hard particles \cite{LDXGalling,NOREM02Friction,SmithThesis}. As such, Tristelle was chosen to be investigated. Tristelle 5183 is an Fe-hardfacing alloy developed from 316 stainless steel. Tristelle 5183 has a similar matrix composition to 316 stainless steel, albeit with the addition of silicon. This means that this investigation can be focused upon the effect of carbides on galling, which, as with the Stellite alloys, appears to result in the improved galling resistance seen for Tristelle 5183 at room temperature when compared with 300-series stainless steels \cite{BowdenThesis,BurdettT5183}. It has been reported that at elevated temperature (\textit{e.g.} \SI{300}{\degreeCelsius}) Tristelle 5183 shows less prominent galling resistance, which has been suggested to occur as a result of a decreased ability to form deformation induced martensite, similar to other Fe-hardfacing alloys \cite{BowdenThesis,KimKim,BurdettT5183,GallingTemperature}.

        \begin{figure}[h]
	    	\centering
	    	\includegraphics[width=8.7cm]{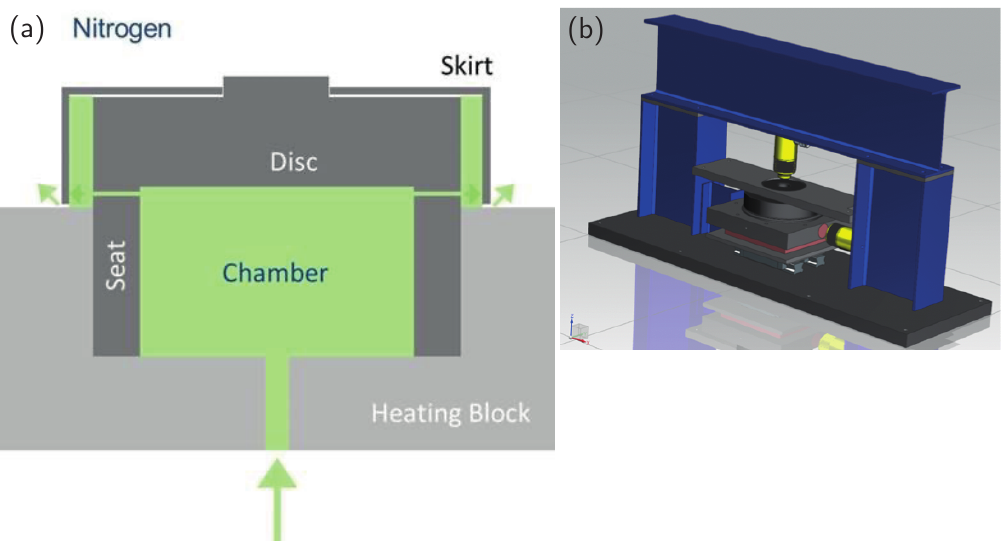}
	    	\caption{Bespoke galling rig manufactured by Rolls-Royce plc. for testing large disk and seat samples. (a) Diagram showing key rig components, (b) CAD drawing showing full rig, including actuators (in yellow). From \cite{LukeSidwellFinalReport}.}
	    	\label{T5183Rig}
	    \end{figure}

        \begin{figure}[h]
	    	\centering
	    	\includegraphics[width=8.7cm]{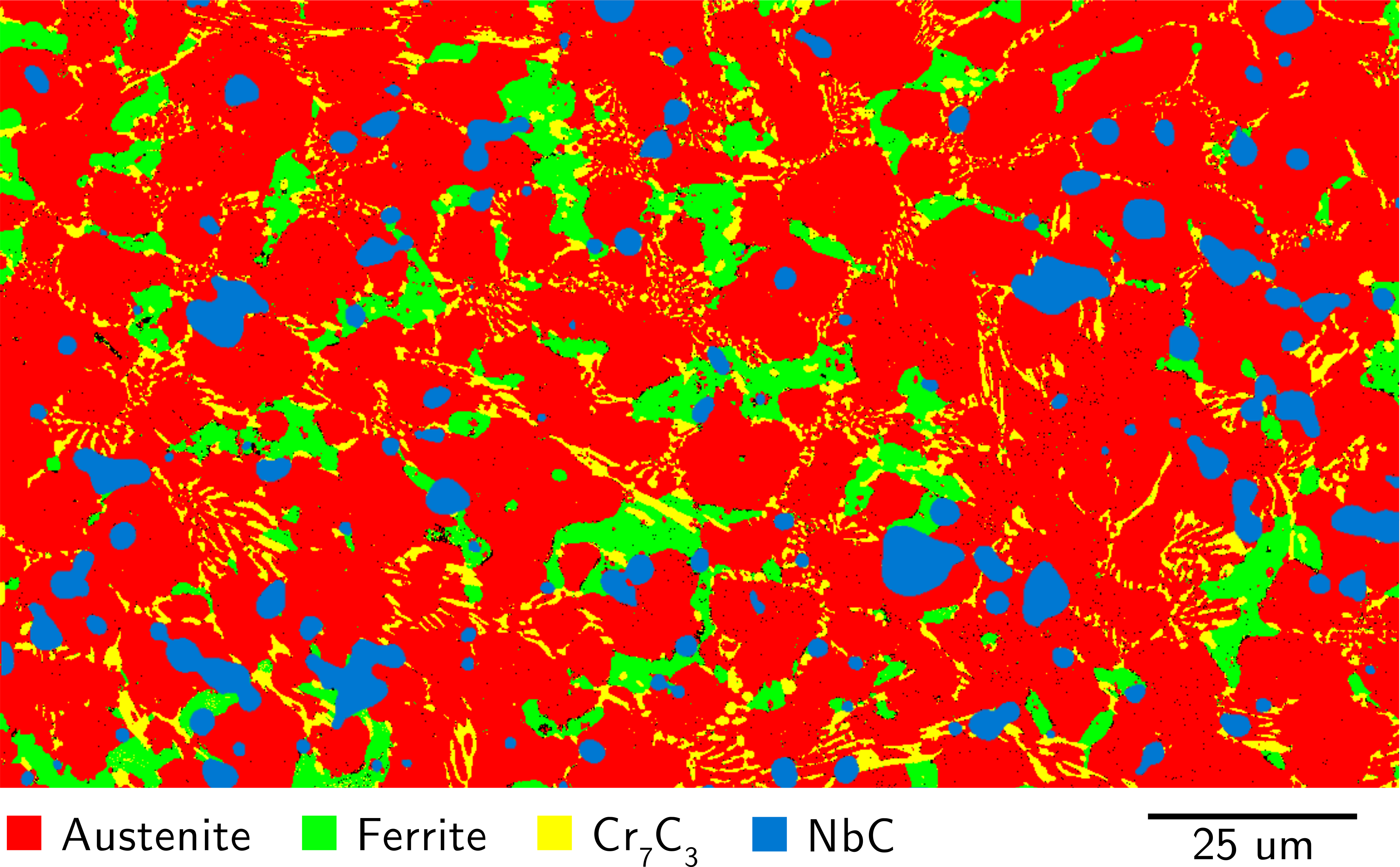}
	    	\caption{Phase map of a Tristelle 5183 weld deposit, created using an EBSD phase map and Nb EDX map.}
	    	\label{T5183EBSD}
	    \end{figure}

   \section{Experimental Methods}
 
    A bespoke rig was manufactured by Rolls-Royce, Figure \ref{T5183Rig}. The disk and seat are made of a 316 stainless steel substrate with a layer of Tristelle 5183 plasma transferred arc (PTA) welded onto its surface and ground down to give an appropriate tribosurface. An actuator was used to apply a normal force to the disc, compressing it against the seat whilst a second actuator was used to slide the seat in a linear manner. The test was stopped early due to an acoustic indication of galling.
        
    Since tests using this rig were conducted at elevated temperatures and oxidation was to be avoided, an inert environment was required. This was created by surrounding the disc with a skirt, and pumping nitrogen into the rig. By applying a positive pressure to the rig, nitrogen can escape the chamber, without air ingress. The test temperature for the test of Tristelle 5183 was approximately \SI{300}{\degreeCelsius}. 
    
	Scanning electron microscopy (SEM) was performed using Zeiss Sigma 300 FEG-SEM, Zeiss Auriga Crossbeam in both secondary electron and backscattered modes, and using a working distance of \SI{7}{}--\SI{10}{\milli\meter} and an accelerating voltage of \SI{20}{\kilo\volt}. An FEI Quanta SEM was used for electron backscattered diffraction (EBSD) and phase maps, using a working distance of \SI{15}{\milli\meter} and an accelerating voltage of \SI{20}{\kilo\volt}. An \textit{in-situ} lift-out was produced using a Helios FIB/SEM to create a transmission electron microscopy (TEM) foil. TEM imaging and diffraction was produced using a JEOL 2100Plus microscope using an accelerating voltage of \SI{200}{\kilo\volt}.
	
	X-ray diffraction was performed using a Bruker D2 Phaser machine, which uses a Cu K-$\alpha$ source with a wavelength of \SI{1.5406}{\angstrom}. An angular resolution of \SI{0.034}{\degree} and time per step of \SI{3680}{\second} were used in this work\footnote{The high time per step is on account of the detector within the Bruker D2 Phaser being able to detect multiple angles at once, in this case 160, meaning that whilst the detector was only stationary in each position for \SI{23}{\second}, the total time a detector was detecting a signal at any given angle was \SI{3680}{\second}.}. Analysis was performed using HighScore Plus software, alongside the ICDD's PDF-2 database.

\begin{table}[t]
\begin{center}
\begin{threeparttable}
\small{
\caption{Tristelle 5183 phase composition in wt.\% and at.\%, as measured using SEM-EDX. C and N are not reliably detectable in EDX and so are not included within the analysis.}
\label{T5183PhaseComp}
\begin{tabular}{lccccccccc}
\toprule
\toprule
\multirow{2}{*}{Phase} & \multicolumn{6}{c}{Composition / wt.\%}                                                                                                            \\
                       & Fe   & Co   & Ni   & Si    &   Cr  &   Nb	 \\
                       \midrule
Austenite		& 61.9 & 2.0  &   12.3   &   3.5    &   20.1   &   0.17 	\\
Ferrite         &   60.6   &   1.7    &   13.0   &   5.1    &   19.2   &   0.40     \\
M$_{7}$C$_{3}$  &   33.5   &   2.8    &   7.5    &   1.6    &   53.8   &   0.83      \\
MC             &   4.2    &   0.2    &   1.1    &   1.0    &   4.5    &   89.0    \\
\bottomrule

\toprule
\multirow{2}{*}{Phase} & \multicolumn{6}{c}{Composition / at.\%}                                                                                                            \\
                       & Fe   & Co   & Ni   & Si    &   Cr  &   Nb	   \\
                       \midrule
Austenite		& 59.5 &   1.8    &   11.2   &   6.7    &   20.7   &   0.10    	\\
Ferrite         &   57.4   &   1.5    &   11.8   &   9.6    &   19.5   &   0.2  \\
M$_{7}$C$_{3}$  &   32.0   &   2.6    &   6.7    & 3.0   &  55.3   &   0.48   \\
MC             &   6.4    &   0.2    &   1.6    &   2.9    &   7.4    &   81.5     \\
\bottomrule
\bottomrule
\end{tabular}}
\end{threeparttable}
\end{center}
\end{table}

\section{Results}
	Prior to investigation of the gall scar, the bulk Tristelle 5183 weld deposit was investigated using EBSD and SEM-EDX, Figure \ref{T5183EBSD}\footnote{Due to austenite and NbC having the same space group, portions of the austenite were often misindexed as NbC. As such, a composite image was created using the EBSD phase map which used austenite, ferrite and Cr$_{7}$C$_{3}$, without NbC, and then overlayed with the Nb EDX map, since Nb was only observed to be present in NbC precipitates.}. As expected, the primary matrix phase is austenite, having formed as dendrites upon cooling. Interdendritic regions contain a eutectic of chromium carbide and ferrite, whilst niobium carbide precipitates are seen to dispersed throughout the microstructure. In the literature, Tristelle 5183 was observed to contain both Cr$_{7}$C$_{3}$ and Cr$_{23}$C$_{6}$, however Cr$_{23}$C$_{6}$ was not observed in this instance. Phase chemistries can be seen in Table \ref{T5183PhaseComp}. The austenite and ferrite chemistries are observed to be very similar, likely as a result of the post-weld solidification and cooling giving insufficient time for diffusion of substitutional solutes to occur, in particular, the Ni and Cr. The niobium carbide precipitates were approximately \textless \SI{0.5}{} -- \SI{10}{\micro\meter} in size, and will have formed first from the liquid, and are thus equiaxed. Chromium carbide precipitates formed in the inter-dendritic regions during final solidification. This results in a variety of sizes and morphologies of chromium carbides being observed.

        \begin{figure}[b!]
	    	\centering
	    	\includegraphics[width=8.7cm]{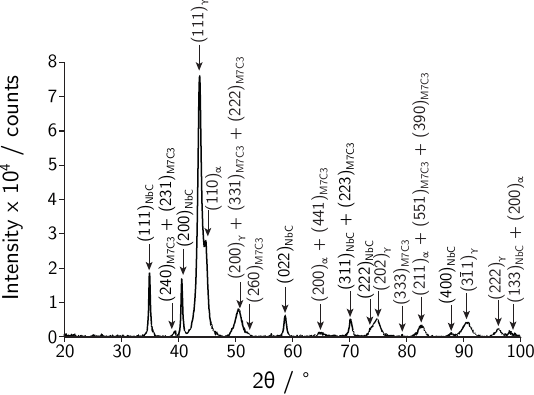}
	    	\caption{X-ray diffraction pattern from a galled Tristelle 5183 surface which was galled at approximately \SI{300}{\degreeCelsius}. Peaks have been labelled, with the exception of minor Cr$_{7}$C$_{3}$ peaks and potential $\alpha$' contributions.}
	    	\label{T5183XRD}
	    \end{figure}
        
         X-ray diffraction (XRD) of the galled Tristelle~5183 surface corroborates the phase results found using EBSD, Figure \ref{T5183XRD}. It is known that austenitic stainless steels may form strain-induced martensite upon deformation\cite{UnOxvsOx}, and since some laths were observed within austenite grains approximately \SI{90}{\micro\meter} below the sample surface, it is reasonable to suggest that martensite has formed. Thus the martensite contributes to the ferrite peaks within the XRD pattern, on account of the martensite having a low tetragonality. In addition, Cr$_{23}$C$_{6}$ is not observed in the XRD pattern, which is in agreement with the EBSD phase data, Figure \ref{T5183EBSD}.

        \begin{figure}[h]
	    	\centering
	    	\includegraphics[width=8.7cm]{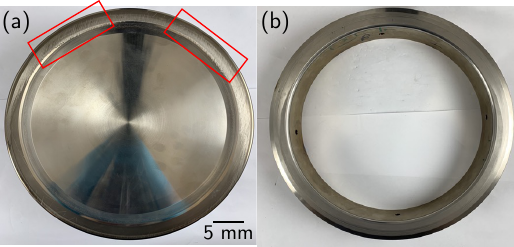}
	    	\caption{Tristelle 5183 (a) disk and (b) seat samples after testing at a temperature of approximately \SI{300}{\degreeCelsius}, however, the normal stress is unknown. Two regions of galling on the disk are highlighted by red boxes. From \cite{LukeSidwellFinalReport}.}
	    	\label{T5183DiskSeat}
	    \end{figure}
    
    Photographs of the disk and seat material can be seen in Figure \ref{T5183DiskSeat}, with two primary areas of galling being visible on the disk, marked with red boxes. These regions were sectioned and subsequently investigated.
    
        \begin{figure}[h]
	    	\centering
	    	\includegraphics[width=8.7cm]{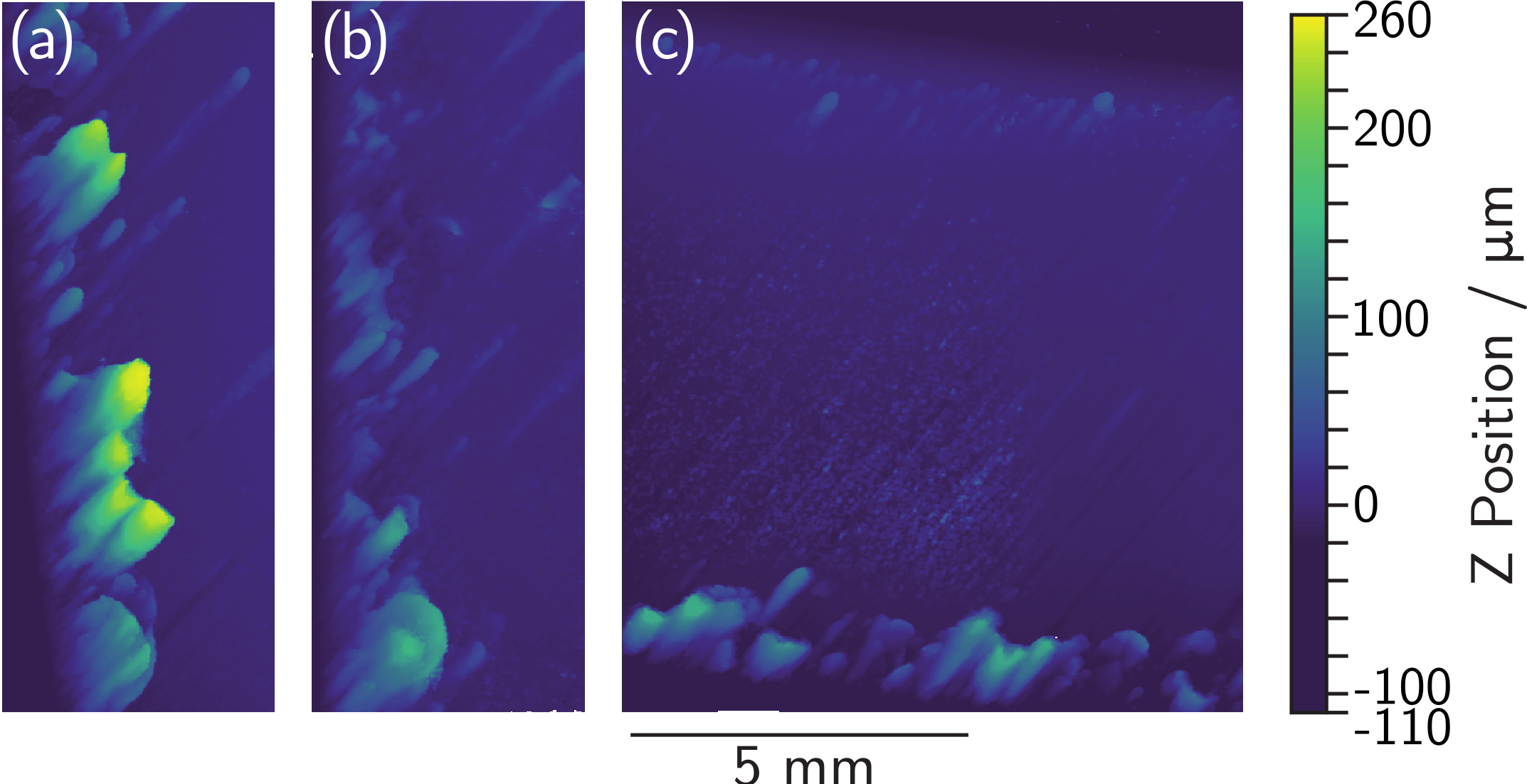}
	    	\caption{White light interferometry images showing small sections of the galled regions highlighted in Figure \ref{T5183DiskSeat}. (a) and (b) show the leading edge of the galled surfaces, whilst (c) shows the full breadth of the galled surface. A common colour scale is used for all images.}
	    	\label{T5183WLI}
	    \end{figure}
    
    Portions of the galling scars shown in Figure \ref{T5183DiskSeat} were investigated using white light interferometry (WLI), Figure \ref{T5183WLI}. Galling peaks are primarily seen to be present from the leading and trailing edges of the samples, with little galling seen in the middle of the tribosurfaces, Figure \ref{T5183WLI}(c). In addition, the galling observed on the trailing edge is considerably reduced in height compared with that seen on the leading edge (maximum heights of \SI{60}{\micro\meter} and \SI{255}{\micro\meter} respectively). Despite showing clear galling peaks, galling troughs were not readily seen in these samples, suggestive of adhesive transfer during galling.
   
         \begin{figure*}[h]
	    	\centering
	    	\includegraphics[width=18.4cm]{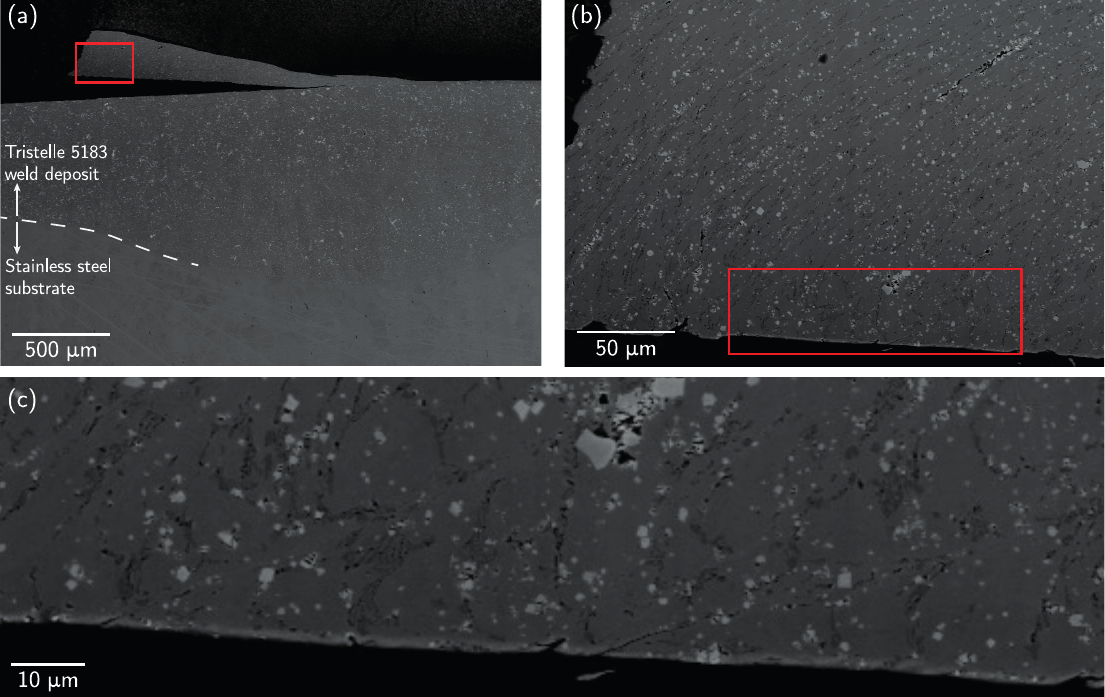}
	    	\caption{A cross-section of a Tristelle 5183 galling peak, parallel to the galling direction. (a) Low magnification, showing the stainless steel substrate under the Tristelle 5183 weld deposit. (b) A higher magnification image, showing a portion of the cast microstructure which has remained undisturbed by the galling process. (c) A higher magnification of the dendrites seen within the galling peak. Due to roughing of the sample after testing, the galling peak has been partially removed from the disk surface.}
	    	\label{T5183PeakDendrite}
	    \end{figure*}
    
    A number of cross-sections parallel to the galling direction were investigated, \textit{e.g.} Figure \ref{T5183PeakDendrite}(a). Due to post-test roughing, the galling peak in Figure \ref{T5183PeakDendrite} has been partially separated from the bulk Tristelle 5183. The stainless steel substrate can also be seen in Figure \ref{T5183PeakDendrite}. Scratches appear across the softer stainless steel substrate as a result of carbide pull-out from Tristelle 5183 during the polishing process, Figure \ref{T5183PeakDendrite}.
    
    Whilst most of the galling peak shows significant material shearing and the presence of a tribologically affected zone (TAZ), Figure \ref{T5183PeakDendrite}(b), a portion of the cast microstructure has remained undisturbed throughout the galling process, observable by the dendrites and unbroken carbides, Figure \ref{T5183PeakDendrite}(c). This is of significance since this indicates that in some cases, material immediately adjacent to the adhered region may not undergo shear. Since the original surface is still visible it can be seen that the galling peak has formed from material which has adhered from the opposing tribosurface (in this case the seat). This is in agreement with the observations made using WLI, Figure \ref{T5183WLI}.
    
        \begin{figure}[h!]
	    	\centering
	    	\includegraphics[width=8.7cm]{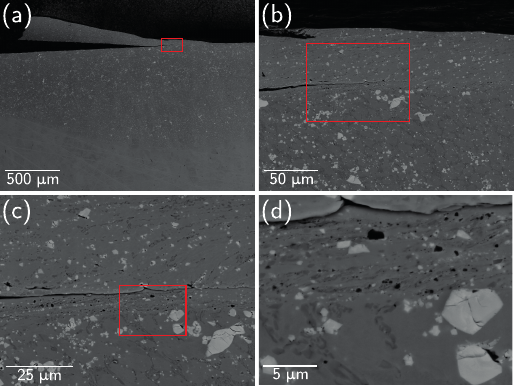}
	    	\caption{A cross-section of the galling peak of Tristelle 5183 shown in Figure \ref{T5183PeakDendrite}, viewing the TAZ in successively higher magnification. Red boxes indicate the region imaged in the following sub-figure.}
	    	\label{T5183TAZ}
	    \end{figure}
	    
        \begin{figure*}[h]
	    	\centering
		\includegraphics[width=18.4cm]{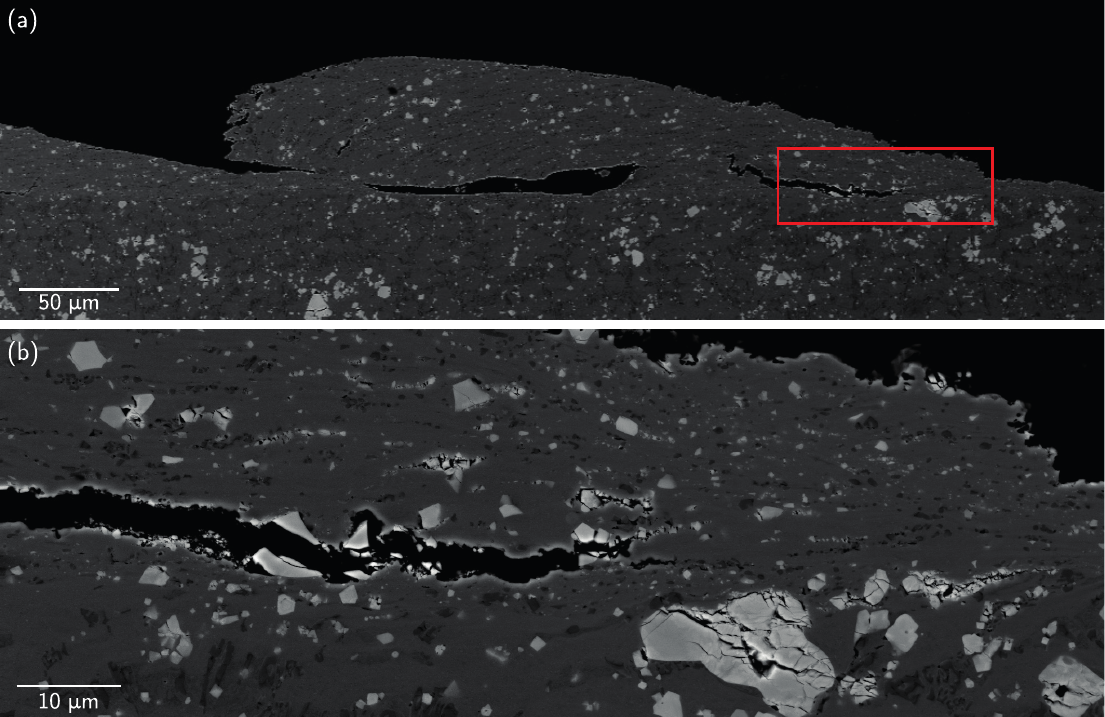}
	    	\caption{A cross-section of a galling peak of Tristelle 5183, viewing the TAZ and crack tip between the adhered galling peak and original surface material. Niobium carbides can be seen to have fractured and when sheared further have the appearance of a stringer. Red box indicates the region imaged in the following sub-figure.}
	    	\label{T5183CarbideFracture}
	    \end{figure*}
	    
	    \begin{figure*}[h!]
	    	\centering
	    	\includegraphics[width=18.4cm]{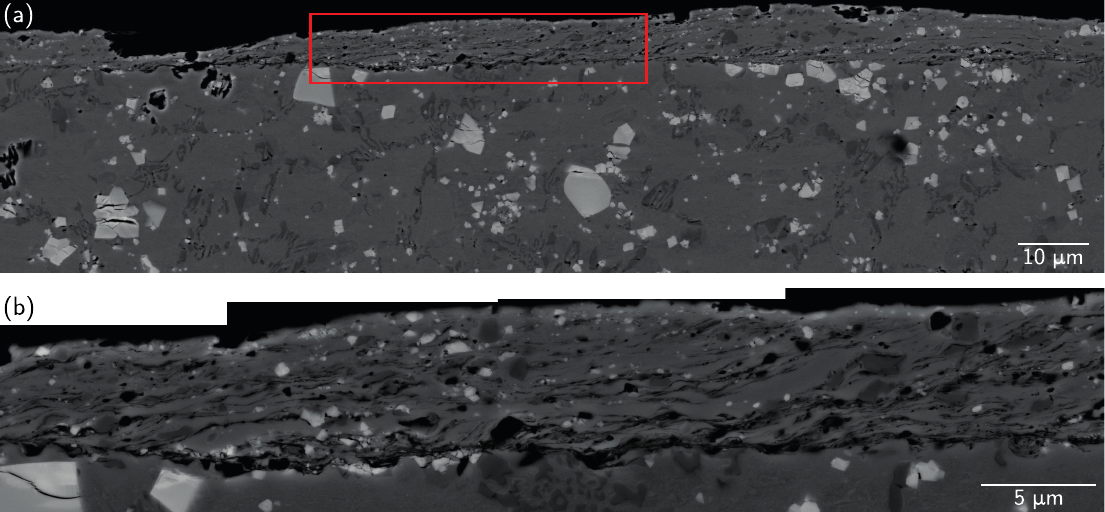}
	    	\caption{A cross-section of a small galling peak of Tristelle 5183 made entirely of the TAZ.}
	    	\label{T5183TAZStitch}
	    \end{figure*}
	    
	    \begin{figure*}[h!]
	    	\centering
	    	\includegraphics[width=18.4cm]{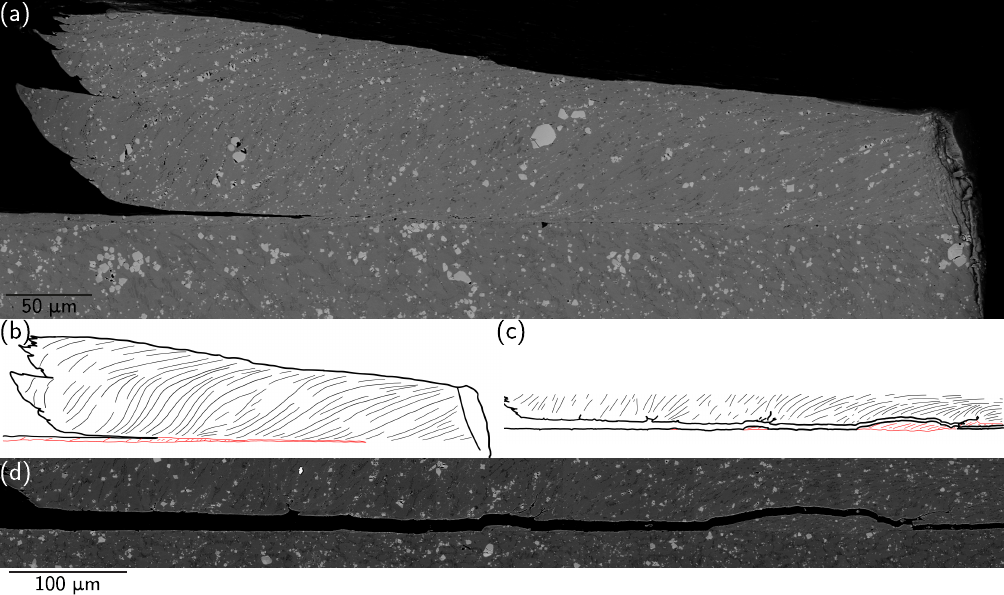}
	    	\caption{Two cross-sections of Tristelle 5183 severed galling peaks: (a) \& (d) viewed in BSE imaging mode and; (b) \& (c) with flow lines shown, and the highly sheared region shown in red. Flow lines were traced from their respective BSE images images.} 

	    	\label{T5183PeakFlow}
	\end{figure*}
	    
    By viewing the crack tip separating the adhered peak seat material from the disk material, it can be seen that the crack has grown at the boundary of two regions within the TAZ with differing amounts of shear, Figure \ref{T5183TAZ}. The more heavily sheared region is approximately \SI{10}{\micro\meter} deep, whilst the less heavily sheared region takes up the remaining material of the galling peak. Although no discernible grain structure is visible within either region, carbide fragments and precipitates show the extent \& direction of shear which are experienced within these regions. This is similar to the way carbide stringers showed the shear direction in 316L stainless steel \cite{UnOxvsOx}.
    
    Another galling peak examined was seen to contain a growing crack, Figure \ref{T5183CarbideFracture}. A number of carbides were observed to have fragmented and been drawn out by shear, resulting in stringer-like formations. Whilst these formations are seen for both chromium- and niobium carbides, voids were only observed to form in the case of fractured niobium carbides. If sufficiently sheared, these voids can coalesce to form a crack, as seen in Figure \ref{T5183CarbideFracture} with carbide fragments on each side of the growing crack.
    
    
    Across the galled regions, a number of smaller peaks can be seen, which are made up of only the TAZ, Figure \ref{T5183TAZStitch}. In this case, there are a significant number of voids which have been formed either from carbide pluck-out close to the TAZ's boundary with the underlying material, or within the TAZ, where carbides were fractured.
    
\begin{figure}[t!]
	    	\centering
	    	\includegraphics[width=8.7cm]{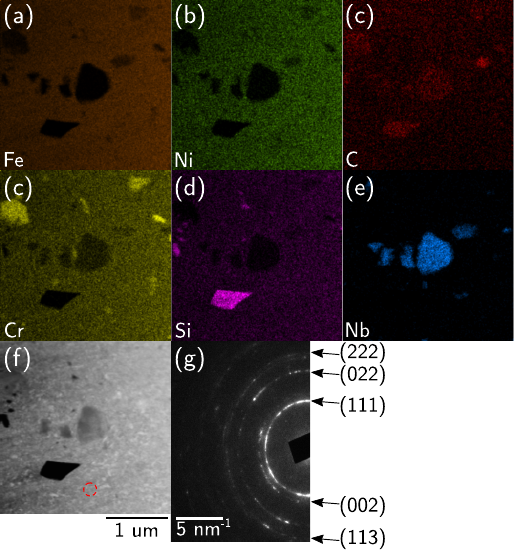}
	    	\caption{(a) - (e) STEM-EDX elemental maps from a TEM foil of Tristelle 5183, taken from a tribologically affected zone (TAZ). (f) An electron image showing the area used to obtain a (g) selected area diffraction pattern, indicated by a dashed red circle.}
	    	\label{T5183STEMEDX}
	\end{figure}

    
    The flow of material within a severed peak and at its adhesion boundary can be seen in Figure \ref{T5183PeakFlow}. Using focussed ion beam milling, a TEM foil of the TAZ was created and analysed, Figure \ref{T5183STEMEDX}. It can be seen that the TAZ contains chromium and niobium carbides which have been significantly reduced in size during galling and its accompanied deformation. In addition, a small isolated SiC particle was observed. The matrix is shown to be nanocrystalline, as evidenced by the almost complete diffraction rings with an aperture of \SI{200}{\nano\meter} and contains only austenite, Figure \ref{T5183STEMEDX}.

\section{Discussion}
    Adhesion and galling is observed to occur between self-mated surfaces of Tristelle 5183 at approximately \SI{300}{\degreeCelsius}. As mentioned previously, adhesive transfer resulted in the formation of galling peaks. Unlike in self-mated tests of 316L stainless steel and non-self mated tests of 316L vs 304L stainless steels, when adhesion occurs, the adhesion junction does not result in trough formation in each tribosurface, Figures \ref{T5183PeakDendrite}, \ref{T5183TAZ} \& \ref{T5183TAZStitch}. In order for galling to occur, one surface must therefore preferentially deform.
    
    \begin{figure}[t]
	    	\centering
	    	\includegraphics[width=8.7cm]{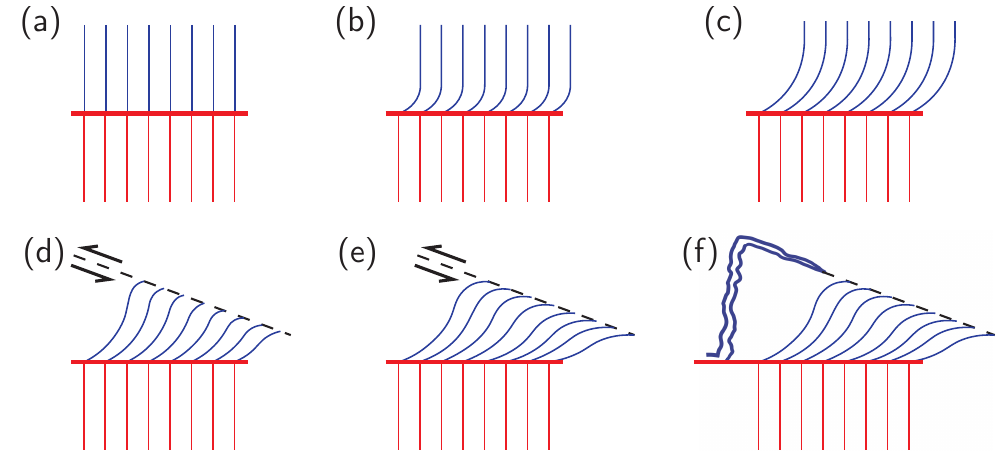}
	    	\caption{The galling mechanism experienced by self-mated Tristelle 5183 at approximately \SI{300}{\degreeCelsius}. (a) An adhesion junction is formed between the two tribosurfaces. (b) Sub-surface shear occurs in only a single tribosurface, and intermittently, as a result of low shear strength regions along the adhesion junction. (c) Shearing continues, resulting in a very heavily sheared region immediately below the adhesion boundary. (d) An internal shear plane is activated. (e) Shear at the adhesion boundary and internal shear boundary continue until (f) fracture occurs, resulting in adhesive transfer.}
	    	\label{T5183Mech}
	    \end{figure}
    
    The galling mechanism experienced by self-mated Tristelle 5183 is shown in Figure \ref{T5183Mech}. Initially, an adhesion junction is formed Figure \ref{T5183Mech}(a), before continued sample movement causes shearing within one of the tribosurface, Figure \ref{T5183Mech}(b). As was noted earlier, shear `shadows' can occur. Shear `shadows' may form due to a lack of adhesion in these locations, whilst adhesion and shear occurs in adjacent regions. As shear continues, a heavily sheared region may form immediately beneath the adhesion boundary, Figure \ref{T5183Mech}(c). If the bulk material is under sufficient shear, an internal shear boundary may form at an angle to the macroscopic direction of movement, Figure \ref{T5183Mech}(d). Eventually, as tribosurface movement continues, Figure \ref{T5183Mech}(e), the shear within the material results in a crack forming and propagating and thus adhesive transfer occurs, Figure \ref{T5183Mech}(f).
    
    This mechanism and the results seen for galled Tristelle 5183 are similar to those seen by Dimaki \textit{et al}.\ who modelled asperities with varying shear strengths, attractive stresses and ductilities, \cite{NatureAdhesionModels}. Tristelle 5183 behaved similarly to a material with a shear stress greater than its attractive stress (whilst the attractive stress was not equal to zero), and a limited ductility. Since the primarily austenitic matrix of Tristelle 5183 is based on 316 stainless steel, it is expected that the matrix has a propensity for adhesion or attraction \cite{UnOxvsOx}. Similarly, due to the incorporation of carbides into its matrix, Tristelle is known to have limited ductility. This was seen in Figure \ref{T5183CarbideFracture}, where carbides were observed to have fractured and formed voids, resulting in the formation and propagation of a crack through void coalescence.  

    In some cases, as in Figure \ref{T5183TAZStitch}, failure occurs at the boundary of the heavily sheared region with the underlying sheared material. This could also have been as a result of void coalescence, or may have occurred due to the change in shear gradients at this point.
    
    On this view of galling, the function of the ceramic phases in galling is primarily to reduce the continuity and extent of the initial adhesion junction, and the junction strength, as ceramics adhere less easily than metals.  The fragmentation of the carbides during the multi-length-scale shear process occurring in the triboilogically affected zone (TAZ) and then during the formation of the galling peaks and troughs presumably limits the shear ductility of the material and promotes earlier, and easier, void formation and hence, cracking and adhesive material transfer. However, even in a nominally single phase material like 316 stainless steel that lacks the ceramic phases in Tristelle 5183 or other hardfacings such as the Stellites \cite{GallingTorque}, very similar overall galling behaviour is observed.

\section{Conclusion}
    Tristelle 5183 is an Fe-based hardfacing alloy containing niobium and chromium carbides within a primarily austenitic matrix (although some ferrite is also present).
    
    Galling and adhesive transfer was found to occur in Tristelle 5183 that was tested at approximately \SI{300}{\degreeCelsius} on a bespoke rig created to closely resemble an in-service valve. Shear at the adhesion boundary on one tribosurface and the subsequent activation of an internal shear plane within this tribosurface resulted in adhesive transfer and galling. The ductility of Tristelle is limited by the incorporation of carbides into the microstructure, which fracture upon sufficient shear, resulting in stringer-like formations, and in the case of niobium carbides, resulted in void formation, coalescence, internal fracture, and subsequent adhesive transfer and galling occurring.
    
    A tribologically affected zone (TAZ) is found to form during galling, which results in the breakup of prior austenite grains into nancrystalline-sized grains in addition to containing fragmented NbC and Cr$_{7}$C$_{3}$.

\section*{Acknowledgements}
We gratefully acknowledge support from Rolls-Royce plc, from EPSRC (EP/N509486/1) and from the Royal Society (D Dye Industry Fellowship).

\bibliography{Galling(All)}
	
\end{document}